\documentclass[10pt,conference]{IEEEtran}
\IEEEoverridecommandlockouts
\usepackage{cite}
\usepackage{amsmath,amssymb,amsfonts}
\usepackage{algorithmic}
\usepackage{graphicx}
\usepackage{textcomp}
\usepackage[usenames,tables]{xcolor}
\usepackage{booktabs}
\usepackage{multirow}
\usepackage{xspace}
\usepackage{xparse}
\usepackage{caption}
\usepackage{subcaption}
\usepackage{footnote}
\usepackage{url}
\usepackage[most]{tcolorbox}
\usepackage{array}
\definecolor{ABlue}{HTML}{127bca}
\definecolor{LHScolor}{HTML}{555555}
\usepackage[hidelinks]{hyperref}

\def\BibTeX{{\rm B\kern-.05em{\sc i\kern-.025em b}\kern-.08em
    T\kern-.1667em\lower.7ex\hbox{E}\kern-.125emX}}
\begin{document}

\font\myfont=ptmb at 13.5pt


\title{\resizebox{\linewidth}{!}{\dataname: Sampling GitHub Issue Report Templates}}


\author{
\IEEEauthorblockN{Nafiseh Nikeghbal$^\star$}
\IEEEauthorblockA{\textit{Sharif University of Technology} \\
nnikeghbal@sharif.edu}
\and
\IEEEauthorblockN{Amir Hossein Kargaran$^\star$}
\IEEEauthorblockA{\textit{CIS, LMU Munich} \\
amir@cis.lmu.de}
\and
\IEEEauthorblockN{Abbas Heydarnoori$^\dagger$}
\IEEEauthorblockA{\textit{Bowling Green State University} \\
aheydar@bgsu.edu}
\and
\IEEEauthorblockN{Hinrich Sch\"utze}
\IEEEauthorblockA{\textit{CIS, LMU Munich} \\
inquiries@cislmu.org}
}

\newcommand\blfootnote[1]{%
  \begingroup
  \renewcommand\thefootnote{}\footnote{#1}%
  \addtocounter{footnote}{-1}%
  \endgroup
}

\newcommand{\ie}{\emph{i.e.,}\xspace}
\newcommand{\eg}{\emph{e.g.,}\xspace}
\newcommand{\etc}{etc.\xspace}
\newcommand{\etal}{\emph{et~al.}\xspace}
\newcommand{\secref}[1]{Section~\ref{#1}\xspace}
\newcommand{\figref}[1]{Fig.~\ref{#1}\xspace}
\newcommand{\listref}[1]{Listing~\ref{#1}\xspace}
\newcommand{\tabref}[1]{Table~\ref{#1}\xspace}
\newcommand{\dataname}{\textsc{GIRT-Data}\xspace}

\newcommand{\allghsnum}{1,106,781\xspace}
\newcommand{\reposoverall}{1,084,300\xspace}
\newcommand{\reposIRTS}{50,032\xspace}
\newcommand{\characteristics}{29\xspace}
\newcommand{\mainlanguages}{19\xspace}

\newcommand{\urltt}[1]{\texttt{\url{#1}}}

\newcommand*\circled[1]{\tikz[baseline=(char.base)]{
		\node[shape=circle,fill,inner sep=0pt] (char) {\textcolor{white}{#1}};}}

\newcommand{\droptextshadow}[2]{%
    \tikz[baseline,outer sep=0pt, inner sep=0pt]{
    \node[#1!40!black] at (0,-0.1ex) {#2};
    \node[white] at (0,0) {#2};
}%
}
\newcommand{\DOIbox}[1]{
\tcbsidebyside[
        bicolor,
        sidebyside,
        sidebyside adapt=both,
        sidebyside gap=5pt,
        top=0pt,left=0pt,right=0pt,bottom=0pt,
        boxrule=0pt,rounded corners,
        interior style={top color=LHScolor,bottom color=LHScolor!60!black},
        segmentation style={top color=ABlue,bottom color=ABlue!60!black},
]{%
\droptextshadow{LHScolor}{DOI}
}{%
\droptextshadow{ABlue}{\href{http://dx.doi.org/#1}{#1}}
}%
}

\newboolean{showcomments}
\setboolean{showcomments}{true}
\ifthenelse{\boolean{showcomments}}
  {\newcommand{\nb}[2]{
    \fbox{\bfseries\sffamily\scriptsize#1}
    {\sf\small$\blacktriangleright$\textit{#2}$\blacktriangleleft$}
   }
  }
  {\newcommand{\nb}[2]{}
  }
\newcommand\EMAD[1]{\textcolor{violet}{\nb{EMAD}{#1}}}
\newcommand\GABRIELE[1]{\textcolor{olive}{\nb{GABRIELE}{#1}}}

\maketitle

\begin{abstract}
GitHub's issue reports provide developers with valuable information that is essential to the evolution of a software development project. Contributors can use these reports to perform software engineering tasks like submitting bugs, requesting features, and collaborating on ideas. In the initial versions of issue reports, there was no standard way of using them. As a result, the quality of issue reports varied widely. To improve the quality of issue reports, GitHub introduced \emph{issue report templates} (IRTs), which pre-fill issue descriptions when a new issue is opened. An IRT usually contains greeting contributors, describing project guidelines, and collecting relevant information. However, despite of effectiveness of this feature which was introduced in 2016, only nearly 5\% of GitHub repositories~(with more than 10 stars) utilize it. There are currently few articles on IRTs, and the available ones only consider a small number of repositories.

In this work, we introduce \dataname, the first and largest dataset of IRTs in both YAML and Markdown format. This dataset and its corresponding open-source crawler tool are intended to support research in this area and to encourage more developers to use IRTs in their repositories. The stable version of the dataset contains \reposoverall repositories and \reposIRTS of them support IRTs.
The stable version of the dataset and crawler is available here: \\\urltt{https://github.com/kargaranamir/girt-data}
\blfootnote{$^\star$ Equal contributions.}
\blfootnote{$^\dagger$ Corresponding author.}
\end{abstract}

\begin{IEEEkeywords}
Issue Report Template, Issue Template, GitHub, Issue Tracker, Bug Report.
\end{IEEEkeywords}

\section{Introduction}\label{sec:intro}
GitHub~\cite{GitHub} hosts over 89 Million public repositories~\cite{GithubCount}, making it the most popular repository hosting service for open-source software projects. Other than hosting Git, GitHub provides several other features, including issue report tracking~\cite{github-issues}. Issue reports address potential software problems, elaborate and discuss code implementation, request feature proposals, collaborate on ideas, track tasks, and work status, support requests and questions, etc. In most popular projects, it is common for there to be tens or hundreds of issues reported each day. Issue reports without an organized structure and essential information complicate issues management activities and increase the developers' workload. Before providing feedback and deciding about the issue, developers need to read and understand the issue description. A well-described issue should reduce the effort required by developers to understand it. However, the quality of issue reports varies widely, and they may need to provide more information to meet developers' needs~\cite{bettenburg2008makes, soltani2020significance}. The result is that developers must ask for more details during subsequent exchanges with contributors (\ie issue openers), increasing the time spent on the discussion.

To overcome this problem, GitHub and other popular hosting platforms for software development (\eg Gitlab) proposed \emph{\textbf{issue report template}} \textbf{(IRT)} feature. IRT allows developers to customize issue structures, including the information contributors are expected to include when opening new issues. IRTs follow either the Markdown or YAML format and are subject to the variables provided by the hosting platform. However, developers can customize IRTs if the rules and variables are respected. IRT usually consists of greeting contributors, explaining the project guidelines, and collecting relevant information~\cite{li2022follow}. Although IRTs were introduced on GitHub in 2016~\cite{github-template-intro} and developers positively rated the usefulness of IRTs on issue reporting~\cite{li2022follow, crystal2021guide, coelho2020github}, they are rarely utilized, according to our analysis.

Only one recent paper~\cite{li2022follow} has empirically analyzed IRTs. This study, however, examined only 802 of the most popular projects.
Therefore, many questions regarding templates still need to be answered on a larger scale. 

\noindent \textbf{Potential Research Questions:}
1) ``What are the attributes of a good IRT?'', 
2) ``How are IRTs distributed across repositories, and how does this distribution vary based on different metrics such as the programming language used?'',
3) ``How has IRT usage evolved over the time?'',
4) ``Do contributors follow the templates?'',
5) ``How to generate an IRT based on the requirements?'', 
6) ``How to evaluate IRTs?'',
7) ``What is the relation of existing IRTs to the project attributes?'',
8) ``How does IRT affect issue tracking, such as issue resolution time and the number of discussions?'',
9) ``What is the impact of IRTs on other studies related to issue reports such as issue report classification or summerization?'',
etc.

The first step in answering such questions is the selection of the subject software repositories and providing a proper dataset of IRTs. GitHub's official APIs~\cite{GitHubAPI} can be used to select subject software repositories and retrieve the data. These APIs, however, are limited in terms of the number of requests that can be generated and the information that can be retrieved. For example, GitHub APIs support up to 1800 authenticated requests per hour for search~\cite{git-rate-search}, and 5000 authenticated requests per hour for all non-search-related requests~\cite{git-rate-non}. The only way to collect data without spending much time is to use some selection criteria, but this is tricky if we do not have a comprehensive overview of the data.

\noindent \textbf{Contributions:} Due to the challenges researchers face researching IRTs, we present \dataname (GitHub IRT Dataset), the first and largest IRT dataset, along with its open-source crawler tool. A stable version of the dataset is hosted on Zenodo~\cite{girtd-dataset}. The crawler behind \dataname can be configured to mine specific projects and update the dataset continuously. \dataname target repositories are selected based on the repositories provided by GHS project~\cite{dabic2021sampling}. As of today, it has gathered information from \reposoverall repositories written in \mainlanguages main different programming languages. IRT studies can be performed efficiently by utilizing the provided dataset to use both the complete set of records or selectively filter and sample repositories based on research needs.

\section{The Dataset}
This section describes \dataname, a dataset from GitHub's \reposoverall public repository containing repository and IRT characteristics. These repositories are selected based on all repositories retrieved by GHS project~\cite{dabic2021sampling}. In GHS, the selected repositories have at least 10 stars. The 10 stars threshold provides a reasonable data quality within the given time constraint and makes the data collection more scalable~\cite{dabic2021sampling}.
IRTs can be found in \reposIRTS of the selected repositories (almost 5\%). Each of these repositories may have more than one IRT. They are all stored in the repository's default branch, in a hidden directory with the path of \texttt{.github/ISSUE\_TEMPLATE}\footnote{We did not consider the legacy IRT Markdown version which is stored in the path of \texttt{.github/ISSUE\_TEMPLATE.md}}. GitHub IRTs can be represented in both YAML and Markdown formats. We found that most of the repositories prefer to use Markdown. The reason may be that configuring IRT in Markdown format is much easier, issue forms follow the Markdown format, and Markdown format gives more flexibility to the contributors. Besides, the other popular hosting platform Gitlab only uses Markdown for IRT. We here gather information for IRTs in both formats. However, since the Markdown format is more popular and less structured than YAML, for the Markdown format, we gathered more characteristics. For repositories with IRTs, both repository and IRT characteristics are provided, and for repositories without IRTs, only repository characteristics are provided.

\textbf{1) Repository Characteristics:} There are \characteristics characteristics (\eg number of stars) associated with each repository, which are stored in the \dataname dataset. Our open-source crawler tool collects these characteristics using the GitHub search API and information on repository landing pages. The \textsc{PyGithub} library~\cite{pygithub-bib} is used to access the GitHub search API, which allows us to collect most of the characteristics. However, some of the characteristics can not be retrieved correctly with GitHub search API, so we used \textsc{XPath Query} to get this information from repository landing pages. For example, the number of contributors cannot always be captured using the GitHub search API. Furthermore, sometimes for some characteristics such as the number of issues, the GitHub search API returns are different from what appears on the repository's issue landing page. The selected repository characteristics can be found at \tabref{tab:repo_characteristics}. According to~\cite{dabic2021sampling}, most of these characteristics are used in previous empirical studies on mining software repositories~\cite{sheoran2014understanding, Han:compsac2019, gonzalez2020state, muse2020prevalence, pecorelli2020developer, 
borrelli2020detecting, bryksin2020using, Bissyande:issre2013, gonzalez2020did, nakamaru2020empirical, Zampetti:saner2019, Coelho:esem18, Tantisuwankul:jss2019}.

\textbf{2) IRT Characteristics:} GitHub IRT in Markdown has a specific format, and it always begins with a table of fields, including \texttt{name}, \texttt{about}, \texttt{assignees}, \texttt{labels}, and \texttt{title}. Following the table is the \texttt{body} of IRT. The issue body allows developers to ensure that contributors provide the necessary information when opening an issue. GitHub IRT in YAML follows a similar but more structured manner, making it much easier to parse and extract information. For GitHub IRT in Markdown, the IRT table and the IRT body fields are stored as IRT characteristics in the \dataname dataset. We provided an anonymized version of the IRT body as a characteristic as well. The anonymized version replaces the personal information (\eg links) with appropriate tokens. However, the GitHub IRT in YAML is just stored as the raw version in \dataname, since it is a more structured document, and by using the \textsc{Pandoc}~\cite{pandoc-bib} library, target details can be extracted when needed. Additionally, YAML can be easily converted to JSON, dataframe, and XML. \tabref{tab:irt_characteristics_md} lists the IRT characteristics and their descriptions for Markdown format, and \tabref{tab:irt_characteristics_yaml} lists those for YAML format.

\begin{table*}[htp]
\centering
\caption{Repository characteristics stored in \dataname for each selected GitHub repository}
\scriptsize
\resizebox{1\textwidth}{!}{
\begin{tabular}{llll}
\toprule
\textbf{Charcteristic} & \textbf{Type} & \textbf{Description} & \textbf{Tool/Method} \\
\midrule

	\texttt{full\_name (key)} & string & Repository full name as \texttt{user\_name/repo\_name} & \textsc{Input} \\
	
	\texttt{has\_IRT} & boolean & Does the IRT path \texttt{.github/ISSUE\_TEMPLATE} exist? & \textsc{PyGithub}\\
	
	\texttt{stargazers\_count} & integer & Number of stars & \textsc{PyGithub}  \\
	
	\texttt{forks\_count} & integer & Number of forks & \textsc{PyGithub} \\
	
	\texttt{subscribers\_count} & integer & Number of users subscribed to get activity notifications (watchers) & \textsc{PyGithub} \\
	
	\texttt{assignees\_count} & integer & Number of issue assignees & \textsc{PyGithub} \\
	
	\texttt{contributors\_count} & integer & Number of contributors & \textsc{XPath Query} \\
	
	\texttt{commits\_count} & integer & Number of commits & \textsc{PyGithub} \\
	
	\texttt{branches\_count} & integer & Number of branches & \textsc{PyGithub} \\
	
	\texttt{releases\_count} & integer & Number of releases & \textsc{PyGithub}\\

	\texttt{last\_modified} & datetime & Latest modification datetime & \textsc{PyGithub} \\
	
	\texttt{pushed\_at} & datetime & Latest push datetime & \textsc{PyGithub} \\

	\texttt{created\_at} & datetime & Repository creation datetime & \textsc{PyGithub}  \\

	\texttt{size} & integer & Size of repository (in kilobytes) & \textsc{PyGithub} \\
	
	\texttt{topics} & list[string] & Topic labels of repository & \textsc{PyGithub} \\

	\texttt{is\_fork} & boolean & Is it a forked repository? & \textsc{PyGithub}\\

    \texttt{has\_wiki} & boolean & Is the repository's wiki enabled? & \textsc{PyGithub} \\

    \texttt{has\_issues} & boolean & Is the repository's issues enabled? & \textsc{PyGithub} \\
	
	\texttt{is\_archive} & boolean & Is the repository archived? & \textsc{PyGithub} \\

	\texttt{main\_language} & string & The main (most used) programming language of repository & \textsc{PyGithub} \\

	\texttt{total\_issues\_count} & integer & Total number of issues (open and closed issues) & \textsc{PyGithub} \\
	
	\texttt{open\_issues\_count} & integer & Number of open issues & \textsc{PyGithub} \\
	
	\texttt{closed\_issues\_count} & integer & Number of closed issues & \textsc{PyGithub} \\

	\texttt{total\_issues\_countv2} & integer & Total number of issues (open and closed) & \textsc{XPath Query} \\
	
	\texttt{open\_issues\_countv2} & integer & Number of open issues & \textsc{XPath Query} \\
	
	\texttt{closed\_issues\_countv2} & integer & Number of closed issues & \textsc{XPath Query} \\

	\texttt{total\_pull\_requests\_count} & integer &Total number of pull requests (open and closed) & \textsc{PyGithub} \\
	
	\texttt{open\_pull\_requests\_count} & integer & Number of open pull requests  & \textsc{PyGithub} \\
	
	\texttt{closed\_pull\_requests\_count} & integer & Number of closed pull requests  & \textsc{PyGithub} \\

\bottomrule
\vspace{0.0pt}
\end{tabular}
}
\label{tab:repo_characteristics}
\end{table*}

\begin{table*}[htp]
\centering
\caption{Characteristics of IRTs in Markdown format stored in \dataname for each downloaded IRT}
\scriptsize
\resizebox{1\textwidth}{!}{
\begin{tabular}{llll}
\toprule
\textbf{Charcteristic} & \textbf{Type} & \textbf{Description} & \textbf{Tool/Method} \\
\midrule

	\texttt{full\_name} & string & Repository full name as \texttt{user\_name/repo\_name} & \textsc{Input} \\
	
	\texttt{IRT\_name} & string & IRT file name & \textsc{PyGithub} \\
	
	\texttt{IRT\_full\_name (key)} & string &  \texttt{full\_name/IRT\_name} & \textsc{Input}, \textsc{PyGithub}  \\
	
	\texttt{has\_initial\_table} & boolean &  Does IRT contain the initial table?  & \textsc{Regex}  \\

	\texttt{name} & string & Name of IRT (unique in the project templates) & \textsc{Regex}  \\
	
	\texttt{about} & string & Description of IRT (displays by template chooser). & \textsc{Regex} \\
	
	\texttt{title} & string & Default title that will be pre-filled in the issue submission form & \textsc{Regex} \\
	
	\texttt{labels} & comma-delimited string & Automatically assigned labels to issues created with this IRT & \textsc{Regex} \\
	
	\texttt{assignees} & comma-delimited string & Automatically assigned users to issues created with this IRT & \textsc{Regex} \\
	
	\texttt{IRT\_raw} & string & Raw (original) version of IRT that has been downloaded in Markdown format & \textsc{PyGithub} \\

	\texttt{body} & string & Main body of IRT & \textsc{Regex} \\
	
	\texttt{body\_anonymized} & string & Anonymized \texttt{body} & \textsc{Pandoc}, \textsc{Regex} \\

	\texttt{headlines} & list[tuple[string, string]] & Emphasis and headlines in order in format of tuple[headline type, headline text] & \textsc{Pandoc} \\

\bottomrule
\vspace{0.0pt}
\end{tabular}
}
\label{tab:irt_characteristics_md}
\end{table*}

\begin{table*}[htp]
\caption{Characteristics of IRTs in YAML format stored in \dataname for each downloaded IRT}
\centering
\scriptsize
\resizebox{1\textwidth}{!}{
\begin{tabular}{llll}
\toprule
\textbf{Charcteristic} & \textbf{Type} & \textbf{Description} & \textbf{Tool/Method} \\
\midrule

	\texttt{full\_name} & string & Repository full name as \texttt{user\_name/repo\_name} & \textsc{Input} \\
	
	\texttt{IRT\_name} & string & Repository full name as \texttt{user\_name/repo\_name} & \textsc{PyGithub} \\
	
	\texttt{IRT\_full\_name (key)} & string &  \texttt{full\_name/IRT\_name} & \textsc{Input}, \textsc{PyGithub}  \\

	\texttt{IRT\_raw} & string & Raw (original) version of IRT that has been downloaded in YAML format  & \textsc{PyGithub}, \textsc{Requests} \\

\bottomrule
\vspace{0.0pt}
\end{tabular}
}
\label{tab:irt_characteristics_yaml}
\end{table*}

\subsection{Data Extraction}

The \dataname data collection process consists of four steps. Each step is explained below:

\textbf{1) Target Repositories:} The target repositories are first set before the dataset is collected/updated. Our selection includes the most recent versions of all repositories in the GHS project \cite{dabic2021sampling}. Based on the selection made on October 26, 2022, \allghsnum unique repositories \texttt{full\_name} were chosen. At the end of the crawl, we ended up with \reposoverall repositories, since some repositories were unavailable to be downloaded or had fewer than 10 stars.

\textbf{2) Target Repository Characteristics:} The crawler behind \dataname can be configured to crawl any repository characteristics as long as this data can be retrieved from the GitHub API search (using \textsc{PyGithub}) and repository landing pages (using \textsc{XPath Query}). We selected \characteristics characteristics for gathering the stable version of \dataname. Most of these characteristics are used as metrics in previous empirical studies~\cite{dabic2021sampling}. In addition to GHS characteristics~\cite{dabic2021sampling}, we also support additional features that could be helpful in future studies, such as  \texttt{has\_IRT}, \texttt{assigneese\_count}, and \texttt{topics}.

\textbf{3) Crawl:} This part collects repository characteristics and downloads IRTs based on a set of selected repositories. The crawler consists of three components because the information of interest needs to be accessed using different methods. Our crawler is authenticated using a token generated by a GitHub user. The crawler can collect/update over $20k$ repositories every day for each authenticated user. Two instances of our crawler were used simultaneously for the first and second halves of the target repositories. Crawler components include:

\begin{enumerate}
\item[a)] The first component checks whether the IRT path \texttt{.github/ISSUE\_TEMPLATE} exist and download all the files in the path. There are two types of files in this directory: Markdown (\texttt{.md}) files and YAML (\texttt{.yaml} or \texttt{.yml}) files. This part uses \textsc{PyGithub} library for locating files and download them. The data for this part is stored as individual IRT files for each repository.

\item[b)] The second component retrieves the repository characteristics that need to be collected using GitHub search API. This part uses \textsc{PyGithub} library.

\item[c)] The third component uses \textsc{XPath Queries} to retrieve repository characteristics from repository landing pages using the \textsc{requests}~\cite{requests-bib} and \textsc{lxml}~\cite{lxml-bib} libraries. The data of components b) and c) are stored as a one single dataframe. In this dataframe, each repository is a row, each characteristic is a column, and repository \texttt{full\_name} is the primary key, see \tabref{tab:repo_characteristics}.
\end{enumerate}

\textbf{4) Target IRT Characteristics:} The data collected by the crawler in component 1) is used to extract IRT characteristics. Besides having the original IRT as \texttt{IRT\_raw} in both YAML and Markdown IRT characteristics, we parse each Markdown IRT and extract characteristics that can help with future studies. For each Markdown IRT, we used a combination of \textsc{Pandoc} and \textsc{Regex}~\cite{regex-bib} libraries to extract table fields, body, and anonymized version of the IRT body as characteristics.
To anonymize, we used appropriate tokens instead of individual personal information. Our tokens are \texttt{<|Image|>} for images, \texttt{<|URL|>} for urls, \texttt{<|Email|>} for emails, \texttt{<|Code|>} for programming codes, and \texttt{<|Repo\_Name|>} for repository name.
Furthermore, \textsc{Pandoc} library is used to extract headlines from each Markdown IRT, which are used to determine the structure of the IRT.
The data of this part is stored as a dataframe. In this dataframe each IRT is a row, each IRT characteristic is a column, and the primary key is \texttt{IRT\_full\_name} which is the composite of repository \texttt{full\_name} and \texttt{IRT\_name}, see \tabref{tab:irt_characteristics_md} and \tabref{tab:irt_characteristics_yaml}.

\subsection{Querying \dataname}
The data collected for all repositories (\tabref{tab:repo_characteristics}) and IRTs (\tabref{tab:irt_characteristics_md}, \tabref{tab:irt_characteristics_yaml}) is stored in tabular, column-oriented format. The data can be imported into a pandas dataframe or a SQL schema. The query function and logical conditions can be used to filter information easily.

\subsection{Preliminary Analysis}
\begin{figure}[t]
    \centering
    \includegraphics[width=0.49\textwidth]{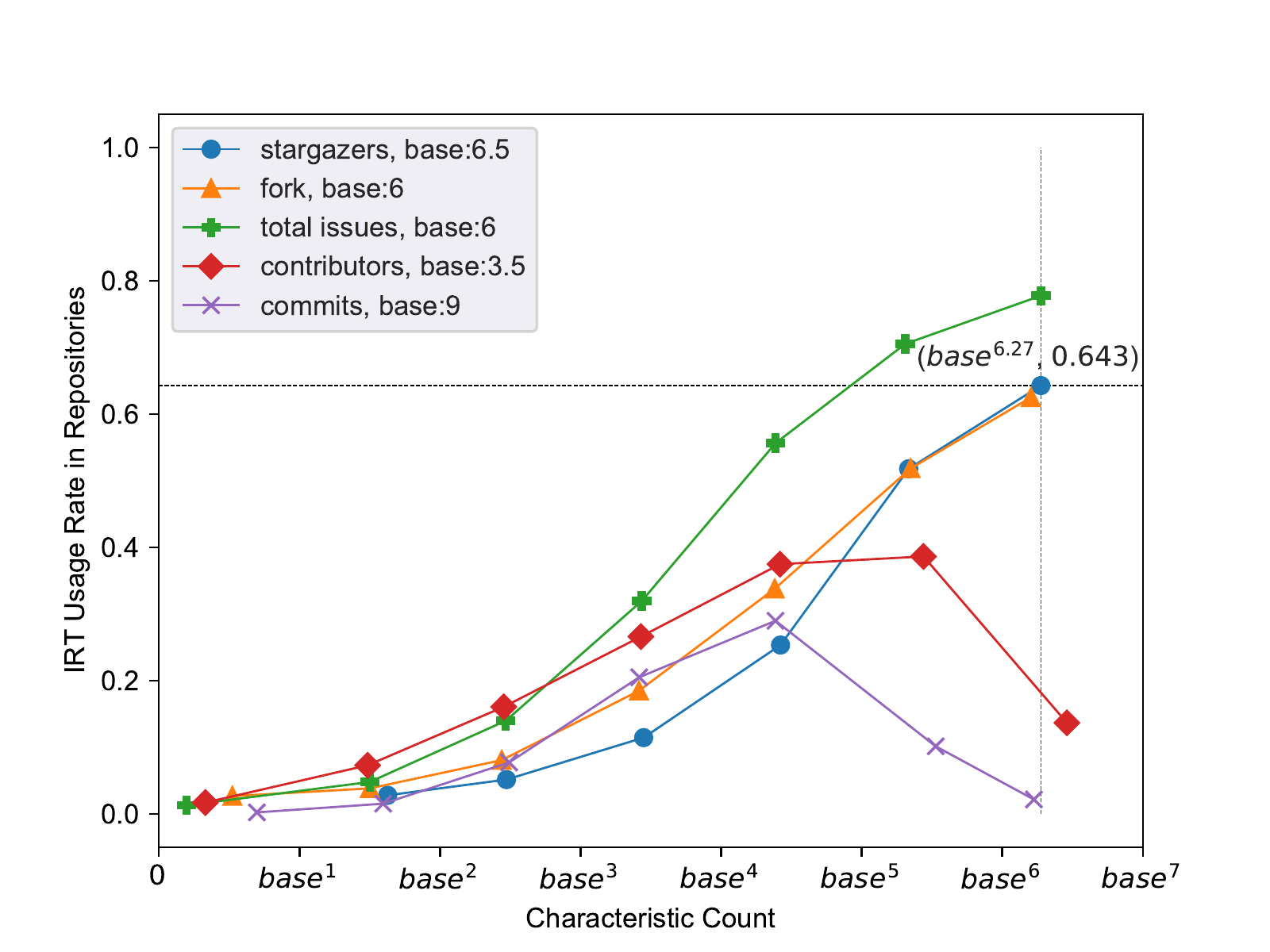}
    \caption{IRT support rates in repositories based on various characteristics. In the case of repositories with a number of stargazers between $6.5^6$ and $6.5^7$~(with an average stargazers number of $6.5^{6.27}$), the odds of supporting an IRT is $0.643$.}
    \label{fig:irt_count_ratio}
\end{figure}

We have preliminarily analyzed the relationship between some repository characteristics and IRT usage rate. \figref{fig:irt_count_ratio} shows the proportion of repositories that use IRTs for each range of repository characteristics. Since the scale of characteristics differs, different axis bases are used for each characteristic. As can be seen, the chance of IRTs being used in repositories increases as most characteristic counts rise. Contributors and commits, however, do not exactly follow this pattern.

\section{Related Work}
\subsection{Support Researchers in MSR}

Several solutions have been proposed to support researchers in MSR in selecting and providing datasets~\cite{dabic2021sampling, GHArchive, Gousi13, dicosmo:hal2017, surana2020tool, markovtsev2018public}. However, none of these studies focus on collecting IRTs while gathering other repository characteristics.

\subsection{Issue Tracking Systems}
There have been several studies conducted to provide effective management of issue reports, such as issue classification and prioritization~\cite{izadi2022predicting, kallis2022nlbse, kallis2021predicting}, issue deduplication~\cite{zhang2022duplicate, messaoud2022duplicate, kim2022predicting}, issue summary generation~\cite{gupta2021approach}, issue title generation~\cite{zhang2022itiger}, and issue structuring~\cite{song2020bee}. However, most of these studies ignore IRTs and their impact on their study.

\subsection{Template Usage in Software Community}
Pull request templates~(PRTs), similar to IRTs, were introduced by GitHub in 2016 to improve pull requests quality. Zhang et al.~\cite{zhang2022consistent} empirically investigated the use of PRTs among 538,864 open-source projects. According to their study, using PRTs positively impacts open-source project maintainability. Li et al.~\cite{li2022follow} do another empirical analysis for IRTs and PRTs. They did the empirical study for 802 of the most popular projects and aimed to find the content, impact, and perception of templates.

To the best of our knowledge, \dataname is the first to comprehensively consider a large scale of IRTs and collect other associated characteristics.

\section{Conclusions and Future Work}
We presented \dataname (GitHub IRT Dataset), a dataset that removes the need to collect and sample repositories for studies on IRTs. The stable version of \dataname containing \reposoverall GitHub repositories written in \mainlanguages main different languages. 
The dataset facilitates the selection of repositories based on diverse criteria for conducting IRT studies.
In the future, we plan to support more repository characteristics, and IRT and PRT characteristics required by the research community.

\bibliography{bibliography}

\begin{thebibliography}{10}
\providecommand{\url}[1]{#1}
\csname url@samestyle\endcsname
\providecommand{\newblock}{\relax}
\providecommand{\bibinfo}[2]{#2}
\providecommand{\BIBentrySTDinterwordspacing}{\spaceskip=0pt\relax}
\providecommand{\BIBentryALTinterwordstretchfactor}{4}
\providecommand{\BIBentryALTinterwordspacing}{\spaceskip=\fontdimen2\font plus
\BIBentryALTinterwordstretchfactor\fontdimen3\font minus
  \fontdimen4\font\relax}
\providecommand{\BIBforeignlanguage}[2]{{%
\expandafter\ifx\csname l@#1\endcsname\relax
\typeout{** WARNING: IEEEtran.bst: No hyphenation pattern has been}%
\typeout{** loaded for the language `#1'. Using the pattern for}%
\typeout{** the default language instead.}%
\else
\language=\csname l@#1\endcsname
\fi
#2}}
\providecommand{\BIBdecl}{\relax}
\BIBdecl

\bibitem{GitHub}
``\url{https://www.github.com},'' 2023.

\bibitem{GithubCount}
``\url{https://api.github.com/search/repositories?q=is:public+fork:true},''
  2023.

\bibitem{github-issues}
``\url{https://docs.github.com/en/issues/tracking-your-work-with-issues/about-issues
  },'' 2023.

\bibitem{bettenburg2008makes}
N.~Bettenburg, S.~Just, A.~Schr{\"o}ter, C.~Weiss, R.~Premraj, and
  T.~Zimmermann, ``What makes a good bug report?'' in \emph{Proceedings of the
  16th ACM SIGSOFT International Symposium on Foundations of software
  engineering}, 2008, pp. 308--318.

\bibitem{soltani2020significance}
M.~Soltani, F.~Hermans, and T.~B{\"a}ck, ``The significance of bug report
  elements,'' \emph{Empirical Software Engineering}, vol.~25, no.~6, pp.
  5255--5294, 2020.

\bibitem{li2022follow}
Z.~Li, Y.~Yu, T.~Wang, Y.~Lei, Y.~Wang, and H.~Wang, ``To follow or not to
  follow: Understanding issue/pull-request templates on github,'' \emph{IEEE
  Transactions on Software Engineering}, 2022.

\bibitem{github-template-intro}
``\url{https://github.blog/2016-02-17-issue-and-pull-request-templates/ },''
  2016.

\bibitem{crystal2021guide}
R.~Crystal-Ornelas, C.~Varadharajan, B.~Bond-Lamberty, K.~Boye, M.~Burrus,
  S.~Cholia, M.~Crow, J.~Damerow, R.~Devarakonda, K.~S. Ely \emph{et~al.}, ``A
  guide to using github for developing and versioning data standards and
  reporting formats,'' \emph{Earth and Space Science}, vol.~8, no.~8, p.
  e2021EA001797, 2021.

\bibitem{coelho2020github}
J.~Coelho, M.~T. Valente, L.~Milen, and L.~L. Silva, ``Is this github project
  maintained? measuring the level of maintenance activity of open-source
  projects,'' \emph{Information and Software Technology}, vol. 122, p. 106274,
  2020.

\bibitem{GitHubAPI}
``\url{https://docs.github.com/en/rest/rate-limit},'' 2023.

\bibitem{git-rate-search}
``\url{https://docs.github.com/en/rest/search?apiVersion=2022-11-28 },'' 2022.

\bibitem{git-rate-non}
``\url{https://docs.github.com/en/rest/overview/resources-in-the-rest-api?apiVersion=2022-11-28
  },'' 2022.

\bibitem{girtd-dataset}
``\url{https://doi.org/10.5281/zenodo.7724792},'' 2023.

\bibitem{dabic2021sampling}
O.~Dabic, E.~Aghajani, and G.~Bavota, ``Sampling projects in github for msr
  studies,'' in \emph{IEEE/ACM 18th International Conference on Mining Software
  Repositories}.\hskip 1em plus 0.5em minus 0.4em\relax IEEE, 2021, pp.
  560--564.

\bibitem{pygithub-bib}
``Pygithub library,'' \url{https://github.com/PyGithub/PyGithub}, 2023.

\bibitem{sheoran2014understanding}
J.~Sheoran, K.~Blincoe, E.~Kalliamvakou, D.~Damian, and J.~Ell, ``Understanding
  ``watchers'' on github,'' in \emph{Proceedings of the 11th working conference
  on mining software repositories}, 2014, pp. 336--339.

\bibitem{Han:compsac2019}
J.~{Han}, S.~{Deng}, X.~{Xia}, D.~{Wang}, and J.~{Yin}, ``Characterization and
  prediction of popular projects on github,'' in \emph{IEEE 43rd Annual
  Computer Software and Applications Conference}, vol.~1, 2019, pp. 21--26.

\bibitem{gonzalez2020state}
D.~Gonzalez, T.~Zimmermann, and N.~Nagappan, ``The state of the ml-universe: 10
  years of artificial intelligence \& machine learning software development on
  github,'' in \emph{Proceedings of the 17th International Conference on Mining
  Software Repositories}, 2020, pp. 431--442.

\bibitem{muse2020prevalence}
B.~A. Muse, M.~M. Rahman, C.~Nagy, A.~Cleve, F.~Khomh, and G.~Antoniol, ``On
  the prevalence, impact, and evolution of sql code smells in data-intensive
  systems,'' in \emph{Proceedings of the 17th International Conference on
  Mining Software Repositories}, 2020, pp. 327--338.

\bibitem{pecorelli2020developer}
F.~Pecorelli, F.~Palomba, F.~Khomh, and A.~De~Lucia, ``Developer-driven code
  smell prioritization,'' in \emph{International Conference on Mining Software
  Repositories}, 2020.

\bibitem{borrelli2020detecting}
A.~Borrelli, V.~Nardone, G.~A. Di~Lucca, G.~Canfora, and M.~Di~Penta,
  ``Detecting video game-specific bad smells in unity projects,'' in
  \emph{Proceedings of the 17th International Conference on Mining Software
  Repositories}, 2020, pp. 198--208.

\bibitem{bryksin2020using}
T.~Bryksin, V.~Petukhov, I.~Alexin, S.~Prikhodko, A.~Shpilman, V.~Kovalenko,
  and N.~Povarov, ``Using large-scale anomaly detection on code to improve
  kotlin compiler,'' \emph{arXiv preprint arXiv:2004.01618}, 2020.

\bibitem{Bissyande:issre2013}
T.~F. {Bissyand\'e}, D.~{Lo}, L.~{Jiang}, L.~{R\'eveill\`ere}, J.~{Klein}, and
  Y.~L. {Traon}, ``Got issues? who cares about it? a large scale investigation
  of issue trackers from github,'' in \emph{IEEE 24th International Symposium
  on Software Reliability Engineering}, 2013, pp. 188--197.

\bibitem{gonzalez2020did}
D.~Gonzalez, M.~Rath, and M.~Mirakhorli, ``Did you remember to test your
  tokens?'' in \emph{Proceedings of the 17th International Conference on Mining
  Software Repositories}, 2020, pp. 232--242.

\bibitem{nakamaru2020empirical}
T.~Nakamaru, T.~Matsunaga, T.~Yamazaki, S.~Akiyama, and S.~Chiba, ``An
  empirical study of method chaining in java,'' in \emph{Proceedings of the
  17th International Conference on Mining Software Repositories}, 2020, pp.
  93--102.

\bibitem{Zampetti:saner2019}
F.~{Zampetti}, G.~{Bavota}, G.~{Canfora}, and M.~D. {Penta}, ``A study on the
  interplay between pull request review and continuous integration builds,'' in
  \emph{IEEE 26th International Conference on Software Analysis, Evolution and
  Reengineering}, 2019, pp. 38--48.

\bibitem{Coelho:esem18}
J.~Coelho, M.~T. Valente, L.~L. Silva, and E.~Shihab, ``Identifying
  unmaintained projects in github,'' in \emph{Proceedings of the 12th ACM/IEEE
  International Symposium on Empirical Software Engineering and
  Measurement}.\hskip 1em plus 0.5em minus 0.4em\relax ACM, 2018.

\bibitem{Tantisuwankul:jss2019}
J.~Tantisuwankul, Y.~S. Nugroho, R.~G. Kula, H.~Hata, A.~Rungsawang,
  P.~Leelaprute, and K.~Matsumoto, ``A topological analysis of communication
  channels for knowledge sharing in contemporary github projects,''
  \emph{Journal of Systems and Software}, vol. 158, p. 110416, 2019.

\bibitem{pandoc-bib}
J.~MacFarlane, A.~Krewinkel, and J.~Rosenthal, ``Pandoc library,''
  \url{https://github.com/jgm/pandoc}, 2023.

\bibitem{requests-bib}
``requests library,'' \url{https://github.com/psf/requests}, 2023.

\bibitem{lxml-bib}
``lxml library,'' \url{https://github.com/lxml/lxml}, 2023.

\bibitem{regex-bib}
``Regular expression library,''
  \url{https://docs.python.org/3/library/re.html}, 2023.

\bibitem{GHArchive}
``\url{https://www.gharchive.org/},'' 2021.

\bibitem{Gousi13}
G.~Gousios, ``The ghtorrent dataset and tool suite,'' in \emph{Proceedings of
  the 10th Working Conference on Mining Software Repositories}.\hskip 1em plus
  0.5em minus 0.4em\relax IEEE Press, 2013, pp. 233--236.

\bibitem{dicosmo:hal2017}
R.~Di~Cosmo and S.~Zacchiroli, ``{Software Heritage: Why and How to Preserve
  Software Source Code},'' in \emph{14th International Conference on Digital
  Preservation}, 2017, pp. 1--10.

\bibitem{surana2020tool}
S.~Surana, S.~Detroja, and S.~Tiwari, ``A tool to extract structured data from
  github,'' \emph{arXiv preprint arXiv:2012.03453}, 2020.

\bibitem{markovtsev2018public}
V.~Markovtsev and W.~Long, ``Public git archive: A big code dataset for all,''
  in \emph{Proceedings of the 15th International Conference on Mining Software
  Repositories}, 2018, pp. 34--37.

\bibitem{izadi2022predicting}
M.~Izadi, K.~Akbari, and A.~Heydarnoori, ``Predicting the objective and
  priority of issue reports in software repositories,'' \emph{Empirical
  Software Engineering}, vol.~27, no.~2, pp. 1--37, 2022.

\bibitem{kallis2022nlbse}
R.~Kallis, O.~Chaparro, A.~Di~Sorbo, and S.~Panichella, ``Nlbse’22 tool
  competition,'' in \emph{IEEE/ACM 1st International Workshop on Natural
  Language-Based Software Engineering}.\hskip 1em plus 0.5em minus 0.4em\relax
  IEEE, 2022, pp. 25--28.

\bibitem{kallis2021predicting}
R.~Kallis, A.~Di~Sorbo, G.~Canfora, and S.~Panichella, ``Predicting issue types
  on github,'' \emph{Science of Computer Programming}, vol. 205, p. 102598,
  2021.

\bibitem{zhang2022duplicate}
T.~Zhang, D.~Han, V.~Vinayakarao, I.~C. Irsan, B.~Xu, F.~Thung, D.~Lo, and
  L.~Jiang, ``Duplicate bug report detection: How far are we?'' \emph{ACM
  Transactions on Software Engineering and Methodology}, 2022.

\bibitem{messaoud2022duplicate}
M.~B. Messaoud, A.~Miladi, I.~Jenhani, M.~W. Mkaouer, and L.~Ghadhab,
  ``Duplicate bug report detection using an attention-based neural language
  model,'' \emph{IEEE Transactions on Reliability}, 2022.

\bibitem{kim2022predicting}
T.~Kim and G.~Yang, ``Predicting duplicate in bug report using topic-based
  duplicate learning with fine tuning-based bert algorithm,'' \emph{IEEE
  Access}, vol.~10, pp. 129\,666--129\,675, 2022.

\bibitem{gupta2021approach}
S.~Gupta and S.~K. Gupta, ``An approach to generate the bug report summaries
  using two-level feature extraction,'' \emph{Expert Systems with
  Applications}, vol. 176, p. 114816, 2021.

\bibitem{zhang2022itiger}
T.~Zhang, I.~C. Irsan, F.~Thung, D.~Han, D.~Lo, and L.~Jiang, ``itiger: an
  automatic issue title generation tool,'' in \emph{Proceedings of the 30th ACM
  Joint European Software Engineering Conference and Symposium on the
  Foundations of Software Engineering}, 2022, pp. 1637--1641.

\bibitem{song2020bee}
Y.~Song and O.~Chaparro, ``Bee: a tool for structuring and analyzing bug
  reports,'' in \emph{Proceedings of the 28th ACM Joint Meeting on European
  Software Engineering Conference and Symposium on the Foundations of Software
  Engineering}, 2020, pp. 1551--1555.

\bibitem{zhang2022consistent}
M.~Zhang, H.~Liu, C.~Chen, Y.~Liu, and S.~Bai, ``Consistent or not? an
  investigation of using pull request template in github,'' \emph{Information
  and Software Technology}, vol. 144, p. 106797, 2022.

\end{thebibliography}
\bibliographystyle{IEEEtran}

\end{document}